\documentclass[10pt]{article}
\usepackage[letterpaper]{geometry}
\usepackage{hicss51}
\usepackage{times}
\usepackage[none]{hyphenat}
\usepackage{url}
\usepackage{latexsym}
\usepackage{indentfirst}
\usepackage{graphicx}
\graphicspath{{images/}}

\usepackage{amssymb,latexsym,amsmath} 
\usepackage{color}

\title{Physics-informed Machine Learning Method for Forecasting and Uncertainty Quantification of Partially Observed and  Unobserved States in Power Grids}

\author{  Ramakrishna Tipireddy \\
  Pacific Northwest National Laboratory \\
 {\underline{ Ramakrishna.Tipireddy@pnnl.gov} } \\\And
  Alexandre Tartakovsky \\
  Pacific Northwest National Laboratory \\
 {\underline{ Alexandre.Tartakovsky@pnnl.gov} }\\\
  \\}

\date{}

\begin{document}
\maketitle

\begin{abstract}

We present a physics-informed Gaussian Process Regression (GPR) model to predict the phase angle, angular speed, and wind mechanical power from a limited number of measurements. In the traditional data-driven GPR method,  the form of the Gaussian Process covariance matrix is assumed and its parameters are found from measurements. In the physics-informed GPR, we treat unknown variables (including  wind speed and mechanical power) as a random process and compute the covariance matrix from the resulting stochastic power grid equations. We demonstrate that the physics-informed GPR method is significantly more accurate than the standard data-driven one for immediate forecasting of generators' angular velocity and phase angle. We also show that the physics-informed GPR provides accurate predictions of the unobserved wind mechanical power, phase angle, or angular velocity when measurements from only one of these variables are available. The immediate forecast of observed variables and predictions of unobserved variables can be used for effectively managing power grids (electricity market clearing, regulation actions) and early detection of abnormal behavior and faults. The physics-based GPR forecast time horizon depends on the combination of input (wind power, load, etc.) correlation time and characteristic (relaxation) time of the power grid and can be extended to short and medium-range times. 

\end{abstract}

\section{Introduction}
Real-time monitoring and immediate forecasting of the power grid dynamics is important for efficient operation of the power grid, including electricity market clearing and regulation actions \cite{soman2010review}.  Other applications that require forecasting include early detection of faults and instabilities  and the effective operation of controllers. Despite the fact that modern power grids are heavily instrumented, directly measuring all  power grid states remains impractical. Inherent high-frequency oscillations of the power grid dynamics, compounded by increasing penetration of renewable energy sources, make immediate forecasting  very challenging. 

Here, we present a ``physics-informed'' Gaussian Process Regression (GPR) method for monitoring (computing unobserved states) and immediate forecasting of power grid dynamics  from measurements of (observed) states. We demonstrate that when partial (past) observations of variables are available, the proposed physics-informed GPR  is significantly more accurate for forecasting than the standard ``data-driven'' GPR. When observations of the variables (states) of interest are not available, the physics-informed GPR still can provide equally good predictions based on measurements of other states, while the data-driven GPR becomes inapplicable.  

 Short-term forecasting in power grids has been addressed with various methods, including machine learning (support vector machines \cite{fentis2016short}, Deep Neural Networks \cite{grant2014short,yeung2017learning,thiyagarajan2016real}) and statistical (e.g.,  the autoregressive integrated moving average method  \cite{nichiforov2017energy}) methods.  GPR \cite{rasmussen2004gaussian}, another machine learning method, has been used for forecasting observed states with the model hyperparameters learned from measurements \cite{blum2013electricity,hachino2014short}. In this work, we refer to this standard GRP method as the ``data-driven'' GPR. Challenges with the data-driven GPR include: 1) it requires a large amount of data collected at high frequency (relative to the process' correlation time), 2) the estimation of hyperparameters can be affected by the noise in measurements, and 3) it cannot predict unobserved states. Unlike some engineered and natural systems, power grid dynamics is well understood, and equations governing power grid dynamics have solid theoretical foundations. Using this characteristic of power grids, we propose a novel physics-informed GPR approach with the covariance matrix computed as a solution of stochastic power grid equations. We consider a grid with wind-powered generators. Because wind mechanical power is oscillating in time and deterministically unknown, it must be modeled as a correlated-in-time random process to properly describe the effect of wind fluctuations on the power grid dynamics and small signal and transient stability \cite
{Rosenthal2017TPS}. Random mechanical wind power transforms power grid equations into stochastic equations \cite{Wang2015JUQ,Barajas2016PRE}.
To compute the covariance matrix, we solve stochastic power grid equation using the strong stability preserving Runge-Kutta (RK) scheme for different realizations of mechanical wind power \cite{milshtein1994numerical,Barajas2016PRE}. 

\section{Problem description}
To introduce the physics-informed GPR approach, we consider a power system with a single wind-powered generator and infinite bus, described by the swing equations \cite{fouad2003power, Wang2015JUQ}:
\begin{align} \label{eqn:dthetadt}
	\frac{d\theta}{dt} &= \omega_B(\omega-\omega_S), \nonumber \\
	\frac{d\omega}{dt} &= \frac{\omega_S}{2H}[P_m-P_e - D(\omega-\omega_S)],
\end{align}
subject to the initial conditions $\theta(t=0) = \theta_0$ and $\omega(t=0) = \omega_0$. 
Here, $\theta \in [0,2\pi)$ is the phase angle between the axis of the generator and the magnetic field, $\omega \in (-\infty,\infty)$ is the angular generator speed, $H$ is the generator inertia, $\omega_B$ is the base speed, $\omega_s$ is the synchronization speed, and $P_e$ is the electric power supplied by the generator to the infinite bus and is related to the phase angle $\theta$ as
\begin{equation} \label{eqn:pe}
	P_e = P_{max} \sin \theta,
\end{equation}
where $P_{max} = \frac{EV}{X}$, $E$ is the internal energy, $V$ is the bus voltage, and $X$ is the total reactance in the system. The mechanical power of wind, $P_m(t) = \langle P_m(t) \rangle + P_m^{\prime}(t)$, is modeled as a stochastic process with mean $\langle P_m \rangle > 0,$ zero-mean fluctuations $\langle P_m^{\prime} \rangle = 0$, variance $\sigma^2$, and the covariance function
\begin{equation}  \label{eqn:rho_exp}
	\langle P_m^{\prime}(t) P_m^{\prime}(s) \rangle = \sigma^2 \exp \left(-\frac{|t-s|}{\lambda} \right),
\end{equation}
where  $\lambda$ is the correlation time. 

To solve the governing equations, we model $P_m^{\prime}(t)$ as an Ornstein-€"Uhlenbeck (O-€"U) process, satisfying the stochastic ordinary differential equation (ODE) \cite{milshtein1994numerical}:
\begin{equation}  \label{eqn:ou_evolve}
	d P_m^{\prime} = -\frac{1}{\lambda} P_m^{\prime} d t + \sqrt{\frac{2}{\lambda}} \sigma dW,
\end{equation}
subject to the initial condition 
\begin{equation}  \label{eqn:ou_evolve}
	P_m^{\prime}(0) = P_{m,0}^{\prime},
	\end{equation}
where $W$ is the standard Wiener process.

The resulting equations are solved using a second-order strong stability preserving RK scheme \cite{milshtein1994numerical} (described in Section \ref{RK}). The initial condition $P_{m,0}^{\prime}(t)$ is obtained from the stationary Gaussian distribution with zero mean and variance $\sigma^2$. The initial conditions for $\theta$ and $\omega$ are set to $\theta_0 = \frac{\arcsin \langle P_m \rangle}{P_{\max}}$ and $\omega_0 = \omega_S$.

\section{Strong stability preserving Runge-Kutta scheme}\label{RK}
To introduce the RK scheme,  we define the vector $y = [\theta, \omega]^T$ and scalar $z = P_m^{\prime}$ functions of time and rewrite the governing equations as
\begin{align} \label{eqn:dthetadt_new3}
	&dy = f(y)dt + g(y) z dt, \nonumber \\	
	&d z = -a z d t + b dW,
\end{align}
where $f(y) = \begin{bmatrix} 
\omega_B(\omega-\omega_S)dt  \\
 \frac{\omega_S}{2H}[ \langle P_m \rangle -P_{\max} \sin \theta - D(\omega-\omega_S)]
\end{bmatrix}$, $g(y) = \begin{bmatrix} 
0  \\
 \frac{\omega_S}{2H} P_m^{\prime} 
\end{bmatrix}$, $a = \frac{1}{\lambda},$ and $b = \sqrt{\frac{2}{\lambda}} \sigma.$ The RK discretization of Eq (\ref{eqn:dthetadt_new3})  is \cite{milshtein1994numerical}: 
\begin{align} \label{eqn:dthetadt_rk}
	&y_{k+1} = y_k + \frac{h}{2} \{(f+gz)_k + (f+gz)_{\bar{k}} \} + \frac{1}{\sqrt{12}} g_k b h^{3/2} \eta_k, \nonumber \\	
	&z_{k+1} = z_k + b \xi_k h^{1/2} +  \frac{h}{2} a (z_k + z_{\bar{k}}) +  \frac{1}{\sqrt{12}} a b h^{3/2} \eta_k,
\end{align}
where $h$ is the time step, $\xi_k$ and $\eta_k$ are the realizations of independent standard Gaussian random variables $\xi$ and $\eta$ at time step $k,$ $f_{\bar{k}} = f(y_{\bar{k}}),$ $g_{\bar{k}} = g(y_{\bar{k}}),$ $y_{\bar{k}} = y_k + (f+gz)_k h$, and $z_{\bar{k}} = z_k +  b \xi_k h^{1/2} + a z_k h.$

\section{Gaussian Process Regression model}\label{sec:GPR}

GPR gives an unbiased estimate of variable  $x^{f} = [x^{f}_1,...,x^f_{N^f}]^T$ ($x^f_i= x^f(t_i)$) at different times $t_i$ and associated uncertainty given observations $x^{o} = [x^{o}_1,...,x^o_{N_o}]^T$ ($x^o_i= x^o(\tilde{t}_i)$).
Here, $x^o$ and $x^f$ can represent the same power grid states, where, for example,  $x^o$ is the observed $\omega$,  $x^f$ is the forecasted $\omega$, and $t_{N_o} < \tilde{t}_1$. Alternatively,   $x^{o}$ and $x^{f}$ could be different states, e.g., $x^{o}$ is observed $\omega$, and $x^f$ is predicted $\theta$ at the same times when $\omega$ is observed (i.e., $t_i=\tilde{t}_i$) or at different times.
    
Let define $x = [x^{o},x^{f}]^T$. Next, we assume that $x$ is the Gaussian process with
\begin{equation}
  \label{eq:joint_gp}
  \begin{bmatrix}
    x^o \\ x^{f}
  \end{bmatrix} \sim \mathcal{N} \left (
    \begin{bmatrix}
      \bar{x}^o \\ \bar{x}^{f}
    \end{bmatrix},
    \begin{bmatrix}
      C_{oo} & C_{of} \\
      C_{fo} &  C_{ff}
    \end{bmatrix}
  \right ),
\end{equation}
where   $\bar{x}^o$ and $\bar{x}^{f}$ are the so-called prior (or, unconditional) mean of $x^o$ and $x^f$, respectively, and $C_{oo}$, $C_{of}$, $C_{fo}$, and $C_{ff}$ are the respective prior covariances between  $x^o$ and $x^o$,  $x^o$ and $x^f$,  $x^f$ and $x^o$, and  $x^f$ and $x^f$. 
Then, GPR defines the conditional (or posterior) estimate of $x^{f}$ given $x^{o}$ as
\begin{equation} \label{eqn:cond_mean}
	\hat{x}^{f}(t) = C_{fo} C_{oo}^{-1}x^o
\end{equation}
and the conditional covariance as 
\begin{equation} \label{eqn:cond_cov}
	\hat{C}_{ff} = C_{ff} - C_{fo} C_{oo}^{-1}C_{of}.
\end{equation}
The main challenge in GPR is to obtain prior statistics, i.e., prior mean and covariance functions. In the standard data-driven GPR method, prior statistics is found from measurements of $x^o$ and  $x^f$ using the so-called marginal likelihood \cite{blum2013electricity,hachino2014short}. Usually, a form of the covariance function is assumed, and the parameters (variance and correlation length) are found by minimizing the negative marginal likelihood of the Gaussian process. Such a purely data-driven approach is impossible to implement for reconstructing unobserved states from observed states without making assumptions  about the correlation between unobserved and observed states. 


Our  proposed physics-informed GPR approach combines GPR  and power grid equations. Specifically, we use Eqs (\ref{eqn:dthetadt}) and (\ref{eqn:ou_evolve}) to compute the mean and covariance functions in Eq (\ref{eq:joint_gp}).  Let $\omega_k^{(n)}$,  $\theta_k^{(n)}$, and $P^{'(n)}_{m,k}$ $k = 0, 1,..., K$ be numerical solutions of Eqs (\ref{eqn:dthetadt}) and (\ref{eqn:ou_evolve}) at time $t_k$ with the initial condition $P^{'(n)}_{m,0}$ ($n = 0, 1, 2, \cdots, N$) drawn from the Gaussian distribution. Then, the mean and covariance of the state $x_k^{(n)}$ $(x=\omega, \theta, P'_m)$ are computed respectively as
\begin{equation}  \label{eqn:mc_mean}
	\bar{x}(t_k) = \bar{x}_k   = \frac{1}{N} \sum_{n=1}^N x_k^n, \quad k = 1, \cdots, K
\end{equation}
and 
\begin{eqnarray}  \label{eqn:mc_cov} \nonumber
	C_{xx}(t_j,t_k)   &=&  \frac{1}{N-1} \sum_{n=1}^N (x_j^n-\bar{x}_j)(x_k^n-\bar{x}_k), \\ 
	&& \quad j,k = 1, \cdots, K.
\end{eqnarray}
The covariance between different variables (e.g., between $\omega$ and $\theta$) is computed as:
\begin{eqnarray}  \label{eqn:mc_cov} \nonumber
	C_{\omega \theta }(t_j,t_k)   &=&  \frac{1}{N-1} \sum_{n=1}^N (\omega_j^n-\bar{\omega}_j)(\theta_k^n-\bar{\theta}_k), \\ 
	&& \quad j,k = 1, \cdots, K.
\end{eqnarray}

\section{Numerical results}

\begin{figure}[h!]
    \centering
        \centering
        \includegraphics[scale=.22]{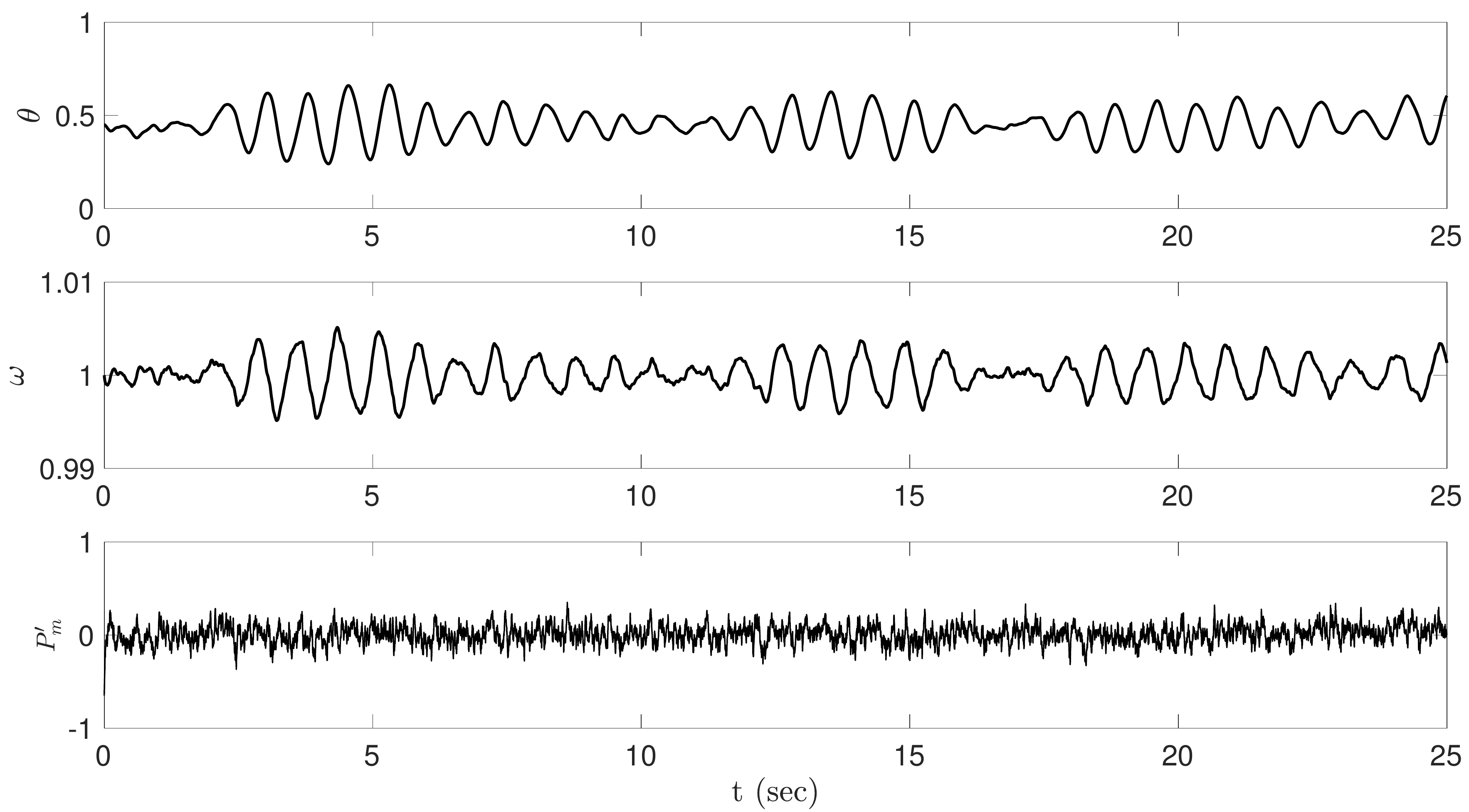}
        \caption{One realization of $\theta(t)$, $\omega(t)$, and $P'_{m}(t)$.} \label{RK:fig:theta_omeag_ou_samle}
\end{figure} 

The numerical solutions are obtained with $P_{\max} = 2.1$ p.u., $H=5$ s, $D=5$ p.u., $\langle P_m \rangle = 0.9$ p.u., $\omega_B=120\pi$ rad/s,  $\lambda = 0.026$ s, and $\omega_S = 1.0.$ The solutions are computed for a total time of $T=25$ s with a time step $h=0.0025$ s. The initial conditions are assumed to be $\theta_0=0.45,$ $\omega_0=1.0$. One thousand Monte Carlo realizations are performed with $P_{m,0}^{'(n)}=\xi^n,$ ($n=0,1,...,999$), where $\xi^n$ is a Gaussian random variable with zero mean and variance $\sigma=0.1$. 

\begin{figure}[h!]
    \centering
        \includegraphics[scale=.2]{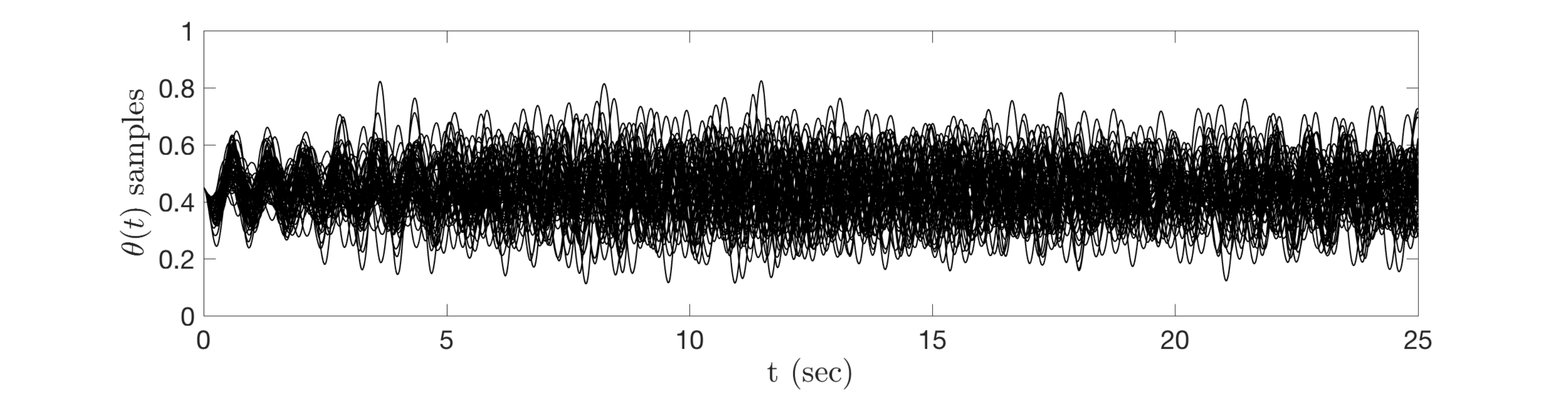}
        \includegraphics[scale=.2]{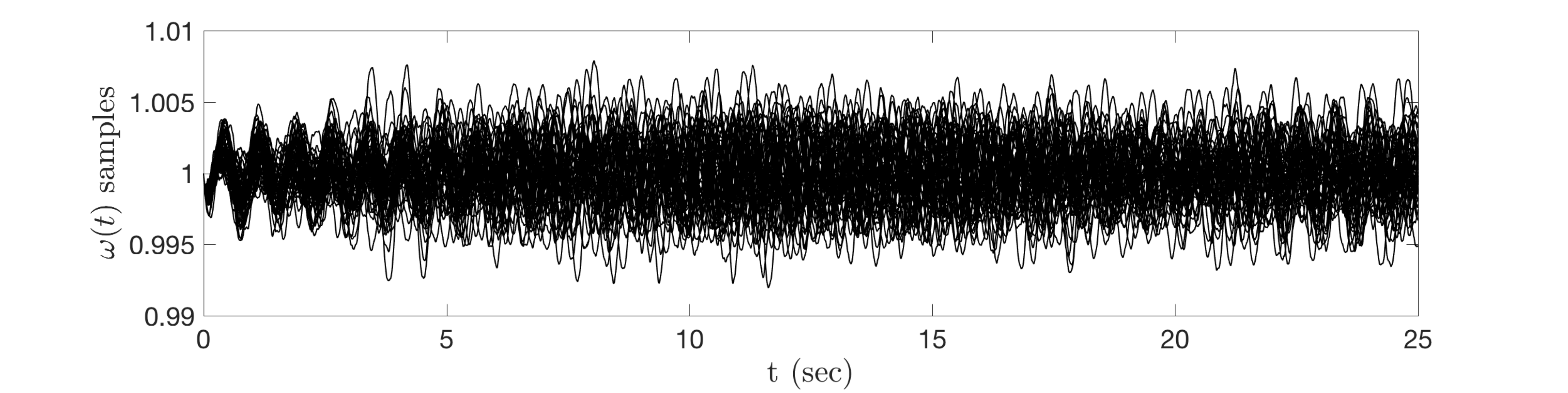}
        \caption{50 realizations of $\theta(t)$ and $\omega(t)$.} \label{RK:fig:omega_samles_50}
\end{figure}    

\begin{figure}[h!]
    \centering
        \includegraphics[scale=.22]{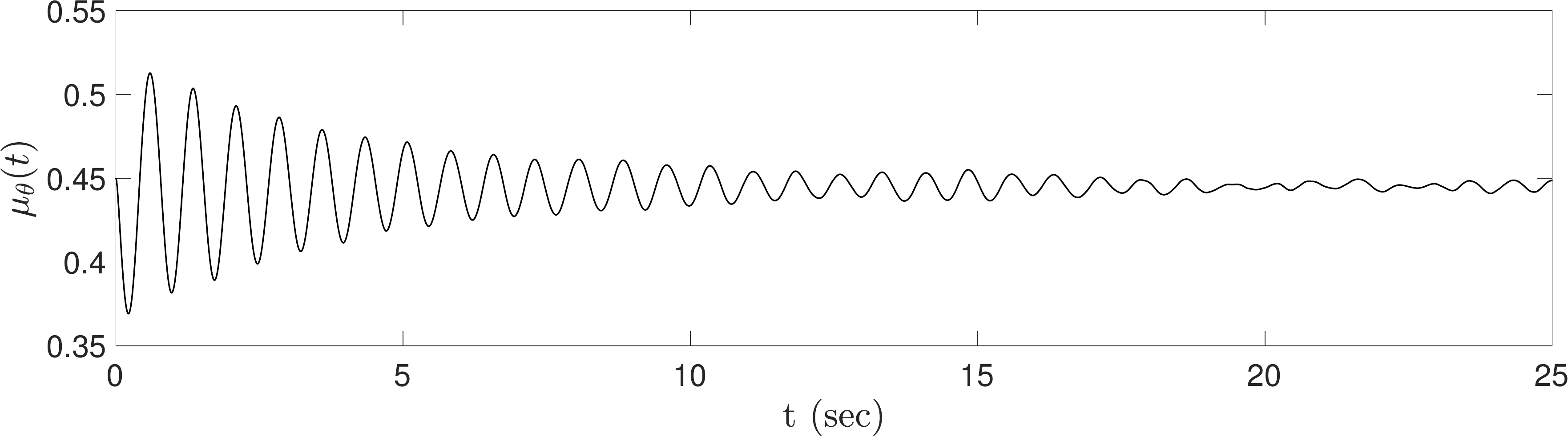}
        \includegraphics[scale=.22]{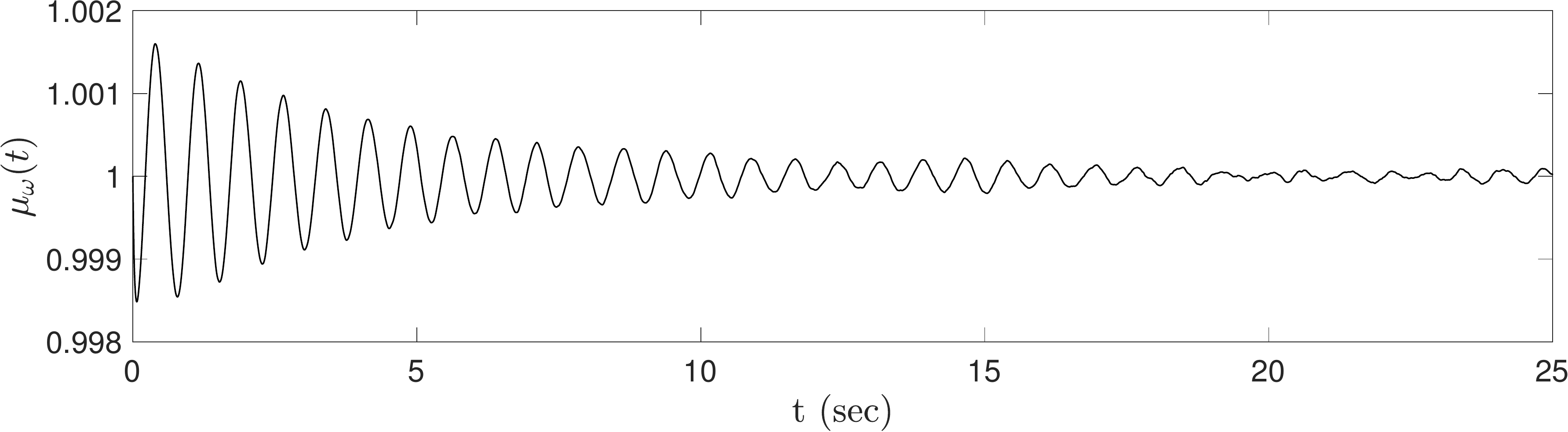}
        \caption{Mean of $\theta$ and $\omega$.} 
        \label{RK:fig:omega_mean_samles_1000}
\end{figure}

\begin{figure}[h!]
    \centering
        \includegraphics[scale=.22]{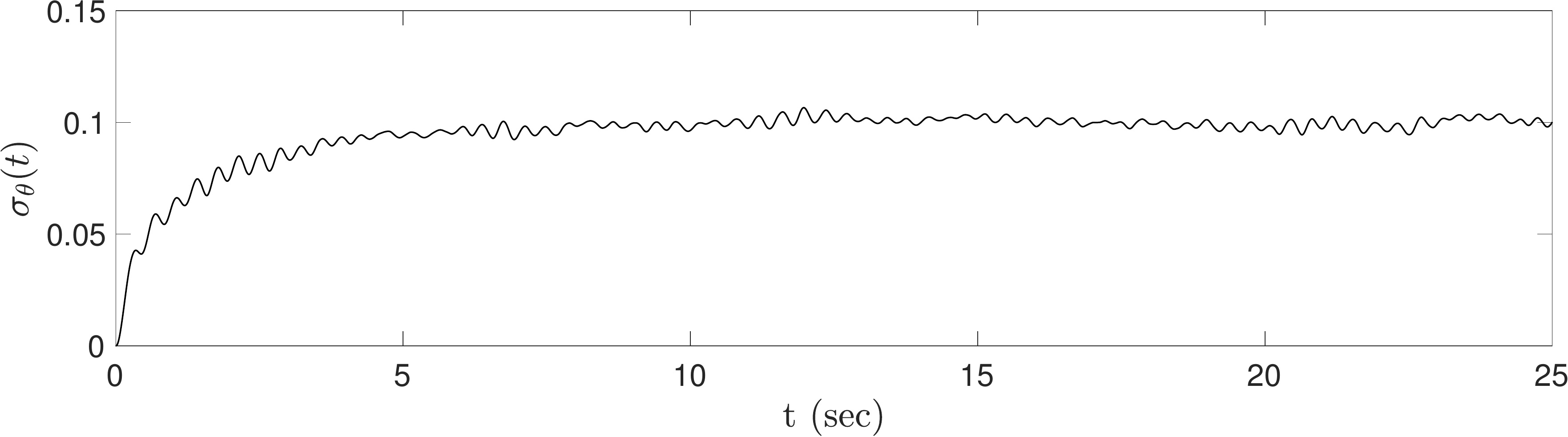}
        \includegraphics[scale=.22]{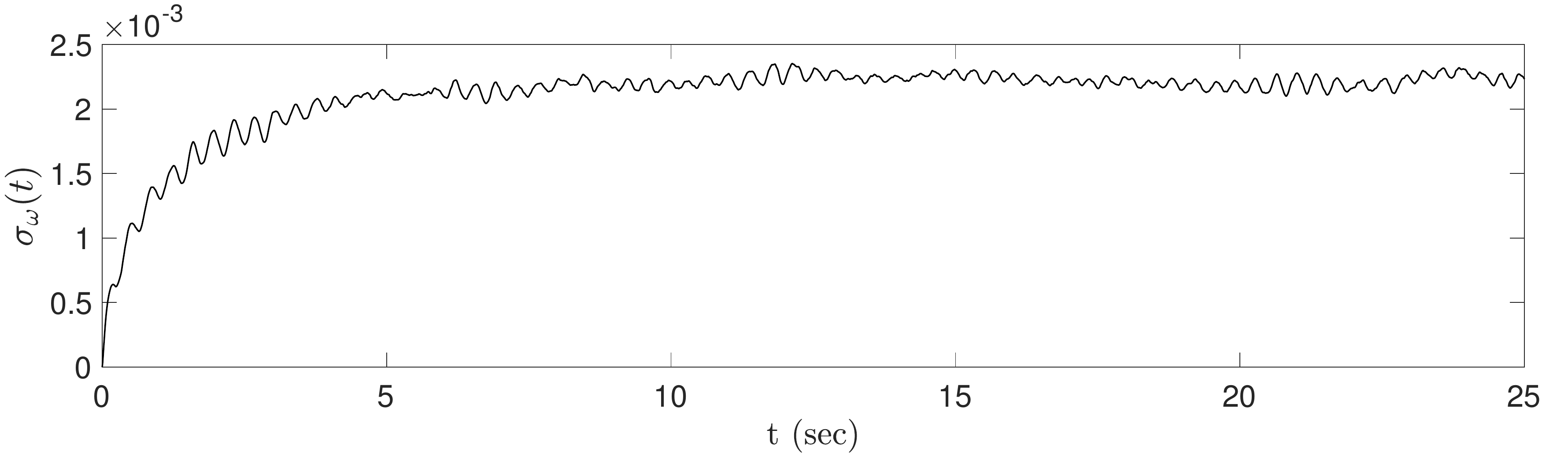}
        \caption{Standard deviation of $\theta$ and $\omega$.} 
        \label{RK:fig:omega_sdev_samles_1000}
\end{figure}

Fig. \ref{RK:fig:theta_omeag_ou_samle} shows one realization of $\theta(t)$, $\omega(t)$, and $P'_{m}(t)$. It illustrates that $\theta$ and $\omega$ have higher-frequency oscillations on the scale of 1 s caused by the high-frequency oscillations in $P'_m$ and lower-frequency oscillations on the scale of 10 s that are proportional to the relaxation time of the system $\frac{H}{\omega_s D}$. In this work, we focus on modeling higher-frequency dynamics.   Fig.   \ref{RK:fig:omega_samles_50} shows 50  realizations of $\theta(t)$ and $\omega(t)$. Mean and standard deviation of $\theta$ and $\omega$  approach asymptotic values after approximately 20 s, which are shown in Figs. \ref{RK:fig:omega_mean_samles_1000} and \ref{RK:fig:omega_sdev_samles_1000}.

We consider three different cases for time series forecasting. In the first case, we assume that partial measurements of $\theta$ and $\omega$ up to  a certain time are available and forecast of $\theta$ and $\omega$ beyond this time using conditional estimation of mean and standard deviation from the Gaussian process model. In the second case, we assume that partial measurements  are available only for $\theta$ and predict $\theta$, $\omega$, and $P'_m.$ In the third case, we predict $\theta$, $\omega$, and $P'_m$ when only partial $\omega$ measurements are available. 

\subsection{Case 1: Forecast of $\theta$ and $\omega$ given past measurements of $\theta$ and $\omega$}

\begin{figure}[h!]
    \centering
        \includegraphics[scale=.22]{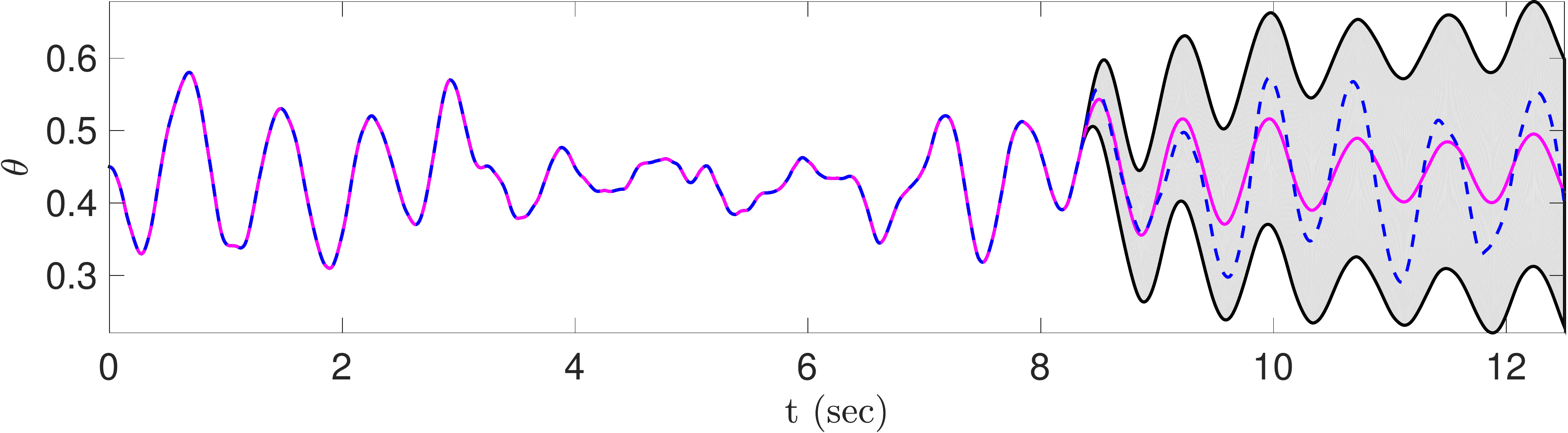}
        \includegraphics[scale=.22]{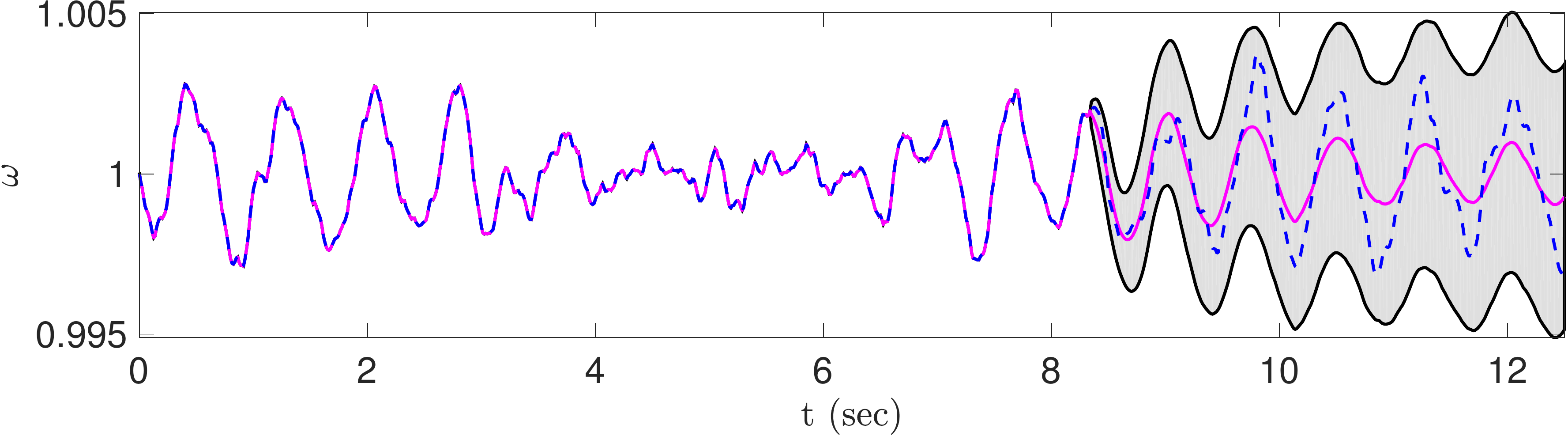}
\caption{Forecast for $\theta$ and $\omega$ (magenta line) and the error bounds (gray area) corresponding to two standard deviations of   $\theta$ and $\omega$, respectively. The blue dashed line shows the ground truth (realization one).   } \label{theta_omega_obs_0}
\end{figure}

In this case, we assume that the $\theta(t)$ and $\omega(t)$ measurements are available until time $t=8.3375$ s every 0.0025 s. Our objective is to forecast $\theta(t)$ and $\omega(t)$ for times greater than $8.3375$ s. To test our approach, we compute prior statistics as described in Section \ref{sec:GPR} and use two randomly selected solutions of the governing equations as ground truth. Specifically, we treat the solution for $t<8.3375$ s  as observations of $\theta(t)$ and $\omega(t)$ and use these observations as input for the GRP model to predict $\theta(t)$ and $\omega(t)$ for $t>8.3375$ s. The second part of the solutions is used to validate the GPR predictions.     

\begin{figure}[h!]
    \centering
        \includegraphics[scale=.22]{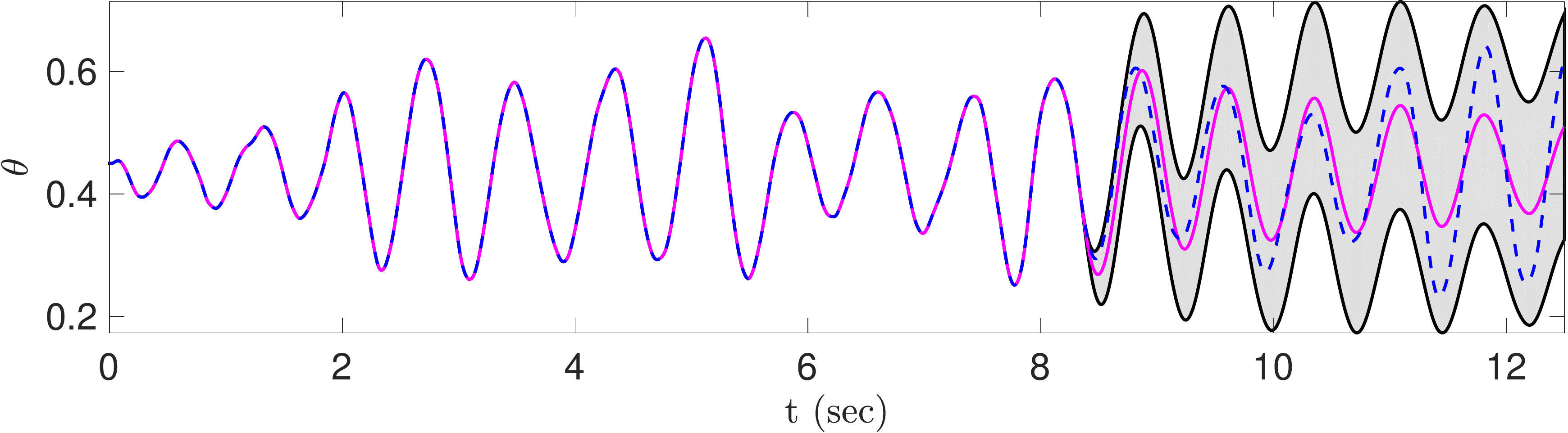}
        \includegraphics[scale=.22]{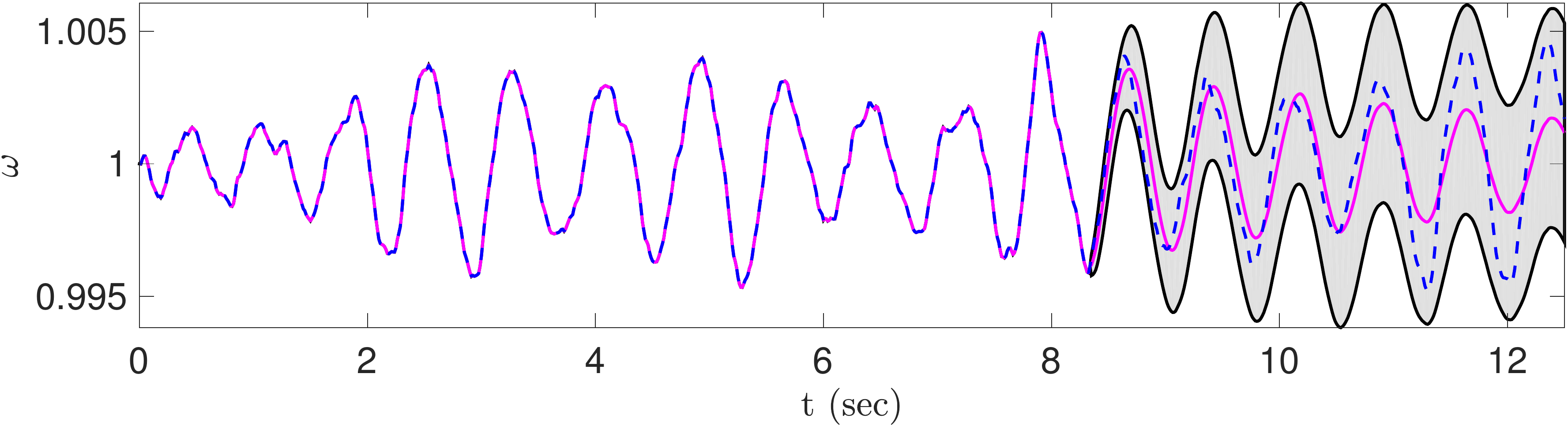}
\caption{Forecast for $\theta$ and $\omega$ (magenta line) and the error bounds (gray area) corresponding to two standard deviations of   $\theta$ and $\omega$, respectively. The blue dashed line shows the ground truth (realization two).} \label{theta_omega_sample_5_obs_0}
\end{figure}

For these two synthetic observation sets, Figs. \ref{theta_omega_obs_0} and \ref{theta_omega_sample_5_obs_0} show the forecast (mean values) of $\theta$ and $\omega$ (magenta line) and uncertainty, represented by  the gray area corresponding to two standard deviations of   $\theta$ and $\omega$. The blue dashed line shows the ground truth. In both cases, the physics-informed GPR method is able to provide an accurate deterministic forecast of $\theta$ and $\omega$ for the time horizon of at least 2 s or 50$\lambda$, i.e., the mean prediction of the GPR closely agrees with the ground truth for at least 20$\lambda$, after which the mean prediction  quality deteriorates.  More importantly, the ``statistical forecast'' is accurate for at least 4 s or 80$\lambda$, i.e., the ground truth lies within the predicted uncertainty envelope.  

\begin{figure}[h!]
    \centering
        \includegraphics[scale=.22]{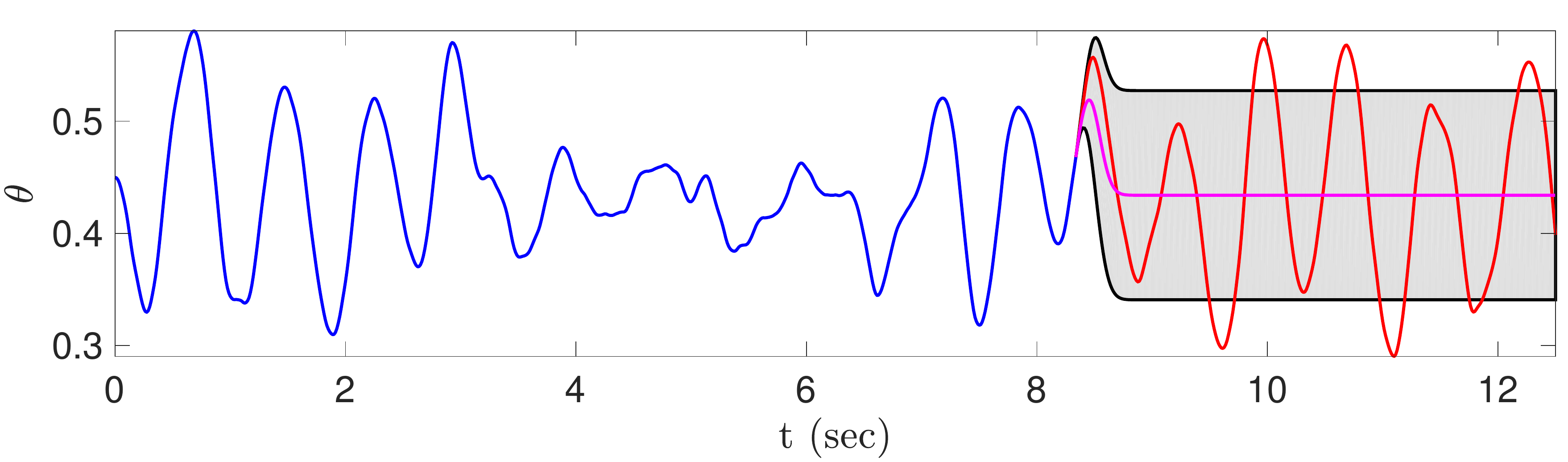}
        \includegraphics[scale=.22]{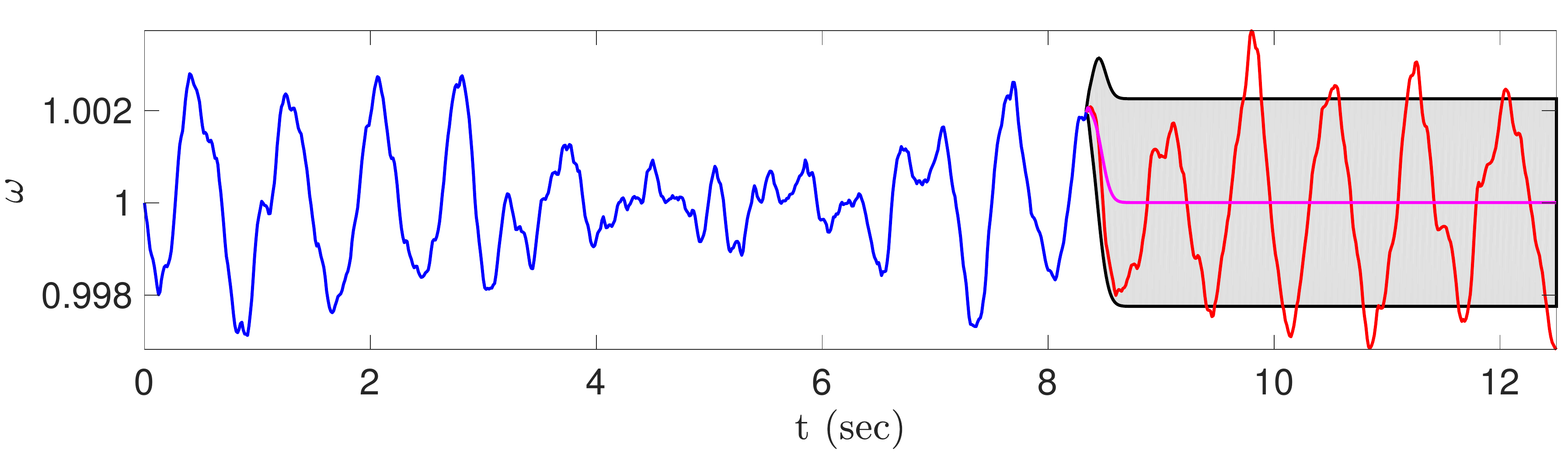}
\caption{Forecast for $\theta$ and $\omega$ (magenta line) and the error bounds (gray area) corresponding to two standard deviations of   $\theta$ and $\omega$, respectively, computed using the standard data-driven GPR with covariance obtained from the marginal likelihood method. Red line shows the ground truth. } \label{theta_omega_standard_GP}
\end{figure}

For comparison, we also forecast $\theta$ and $\omega$ using the standard data-driven GRP method with the covariance obtained from the marginal likelihood method (Fig. \ref{theta_omega_standard_GP}). The comparison of the two methods demonstrates that the the physics-informed GPR performs much better than the data-driven GPR. The data-driven GPR provides an accurate mean prediction for less than 5$\lambda$ and statistical prediction for less than 20$\lambda$. 

\subsection{Case 2: Forecast $\theta$, $\omega$, and $P'_m$ given measurements of $\theta$}
\begin{figure}[h!]
    \centering
        \includegraphics[scale=.22]{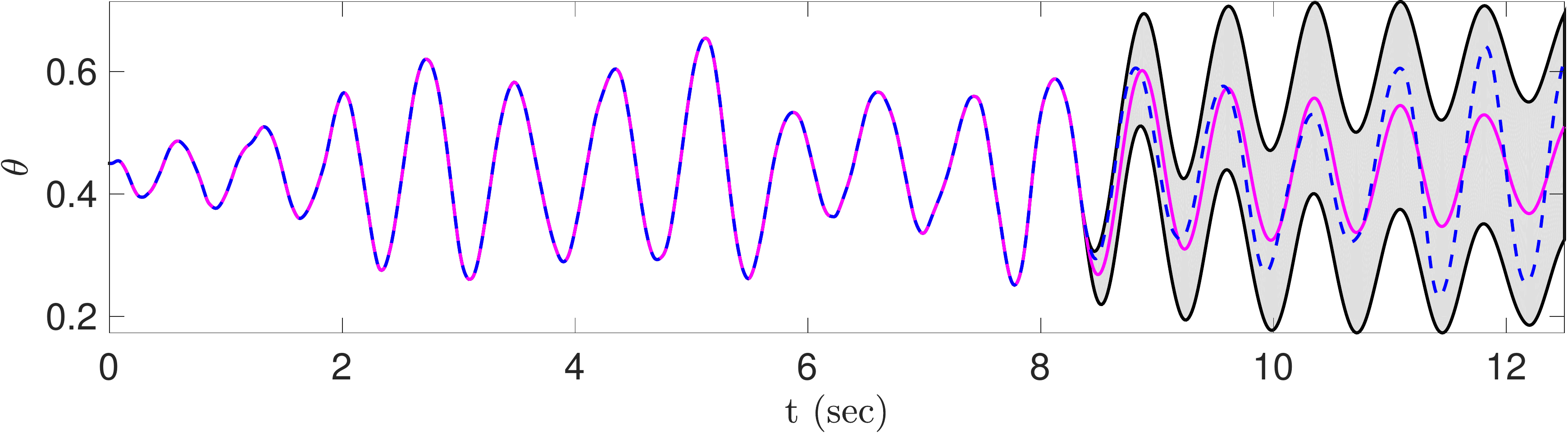}
        \includegraphics[scale=.22]{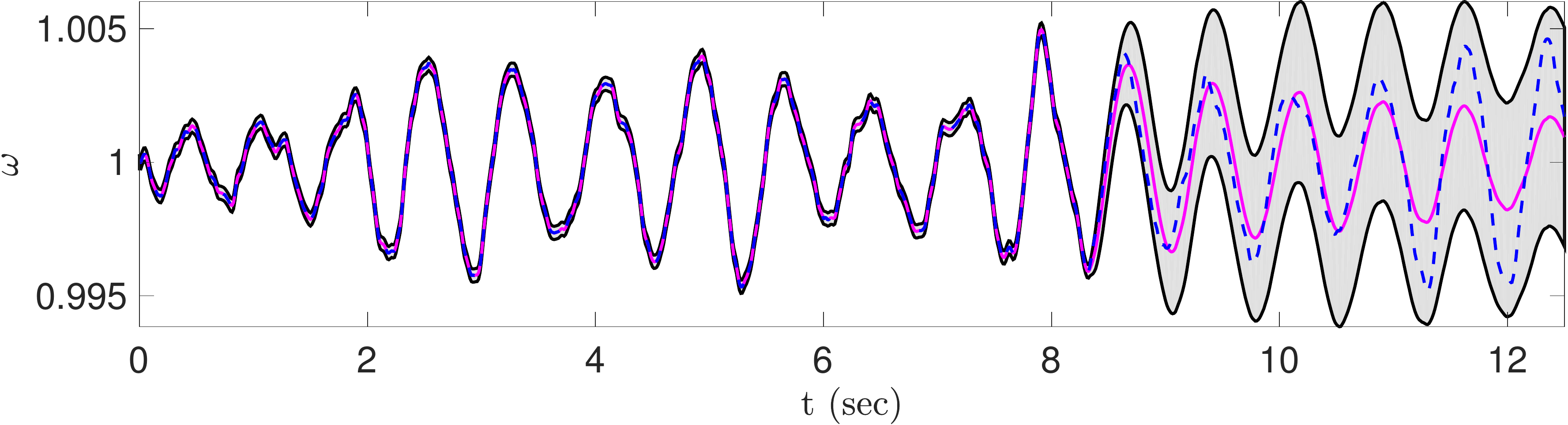}
        \includegraphics[scale=.22]{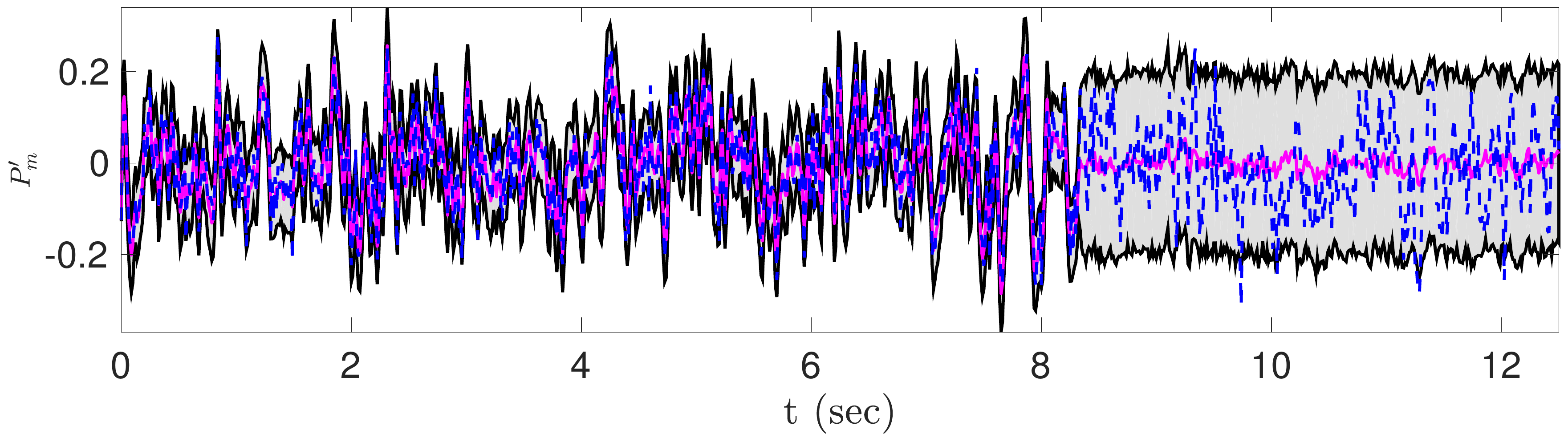}
\caption{Forecast of $\theta$, $\omega$, and $P'_m$ with no  observations of $\theta$ for $t > 8.3375$.} \label{theta_omega_sample_10_obs_theta_0}
\end{figure}

In this case, we assume that   $\theta(t)$ measurements are available for $t<8.3375$ s every 0.0025 s. Our goal is to predict $\theta(t)$ for $t > 8.3375$ and $\omega(t)$ and $P'_m(t)$ for the entire time interval $t\in[0,12.5]$ s. As in Case 1, we compute prior statistics by solving the stochastic power grid equations and use a randomly selected solution as observations of  $\theta(t)$ for $t<8.3375$ s  and ground truth to validate the physics-informed GPR predictions of $\theta(t)$ (for $t > 8.3375$), $\omega(t)$, and $P'_m(t)$. 
Fig. \ref{theta_omega_sample_10_obs_theta_0} shows the predictions of $\theta$, $\omega$, and $P'_m$ and the ground truth results. Notably, the forecast of $\omega$ is exactly the same as in Case 1 (Fig \ref{theta_omega_sample_5_obs_0}) because the $\omega$ observations are the same in these two cases. The prediction for $\theta(t)$ is excellent for $t <8.3375$ when  $\omega(t)$ measurements are available. This is evident by the mean prediction being very close to the ground truth and very small standard deviation (i.e., the narrow gray ``uncertainty'' area). For  $t >8.3375$, the agreement between mean prediction and ground truth  is almost as good as in Case 1. Also, as in Case 1, the ground truth remains within the uncertainty envelope. Similar results are observed for $P_m$. These results demonstrate that the physics-informed GPR is very accurate for computing unobserved variables. This is important because some variables, including $P_m$, are difficult to measure directly.   

\subsection{Case 3: Forecast of $\theta$, $\omega$, and $P'_m$ given measurements of $\omega$}
\begin{figure}[h!]
    \centering
        \includegraphics[scale=.22]{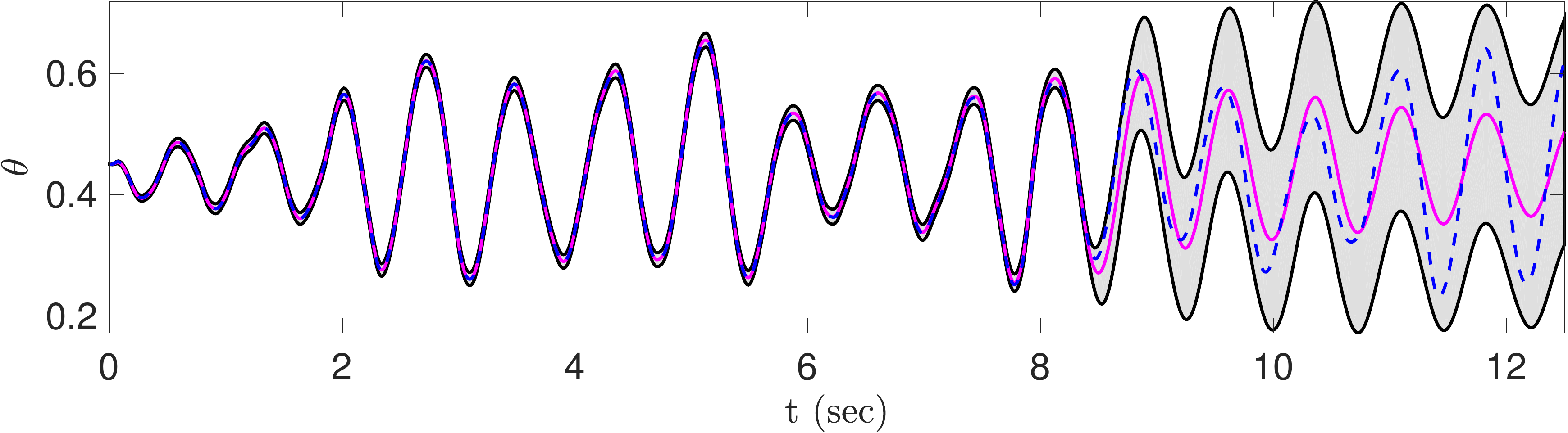}
        \includegraphics[scale=.22]{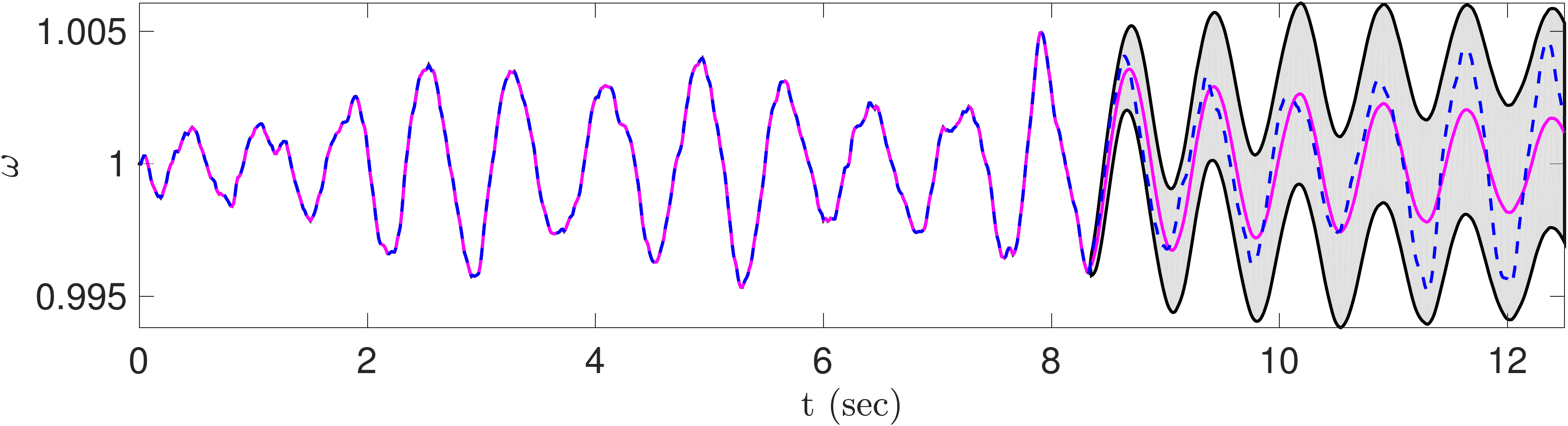}
        \includegraphics[scale=.22]{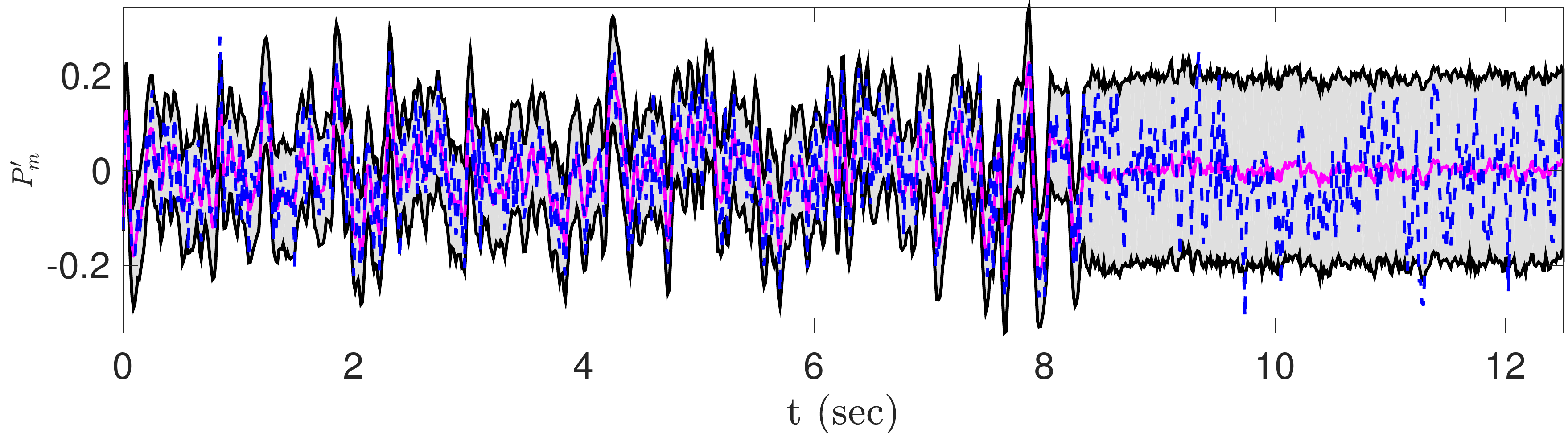}
\caption{Forecast of $\theta$, $\omega$, and $P'_m$ process with no  observations of $\omega$ for $t > 8.3375$. } \label{theta_omega_sample_10_obs_omega_0}
\end{figure}

\begin{figure}[h!]
    \centering
        \includegraphics[scale=.22]{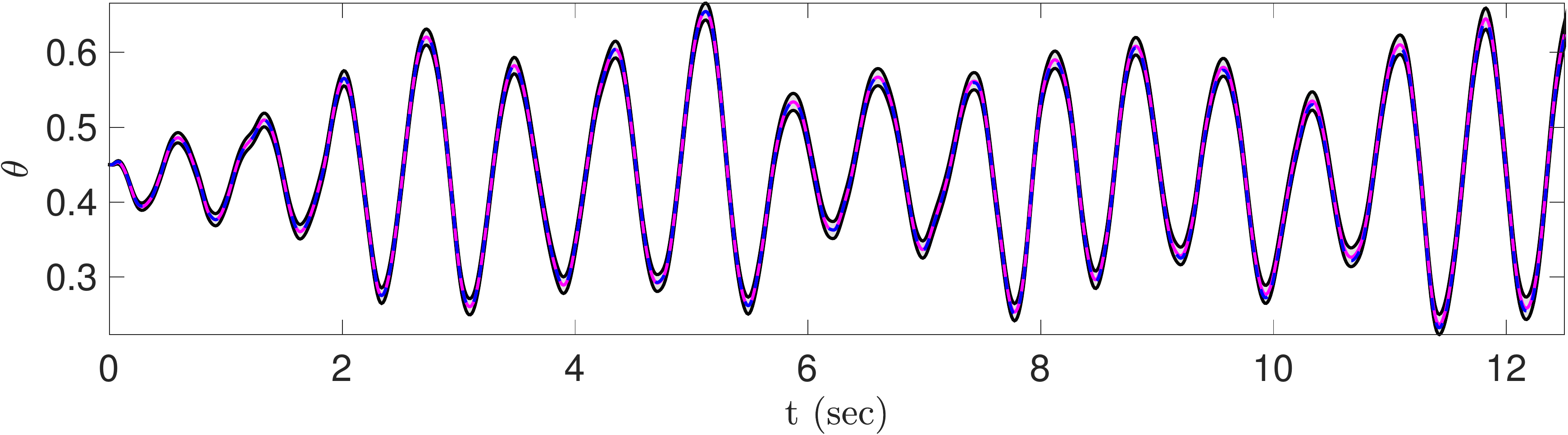}
        \includegraphics[scale=.22]{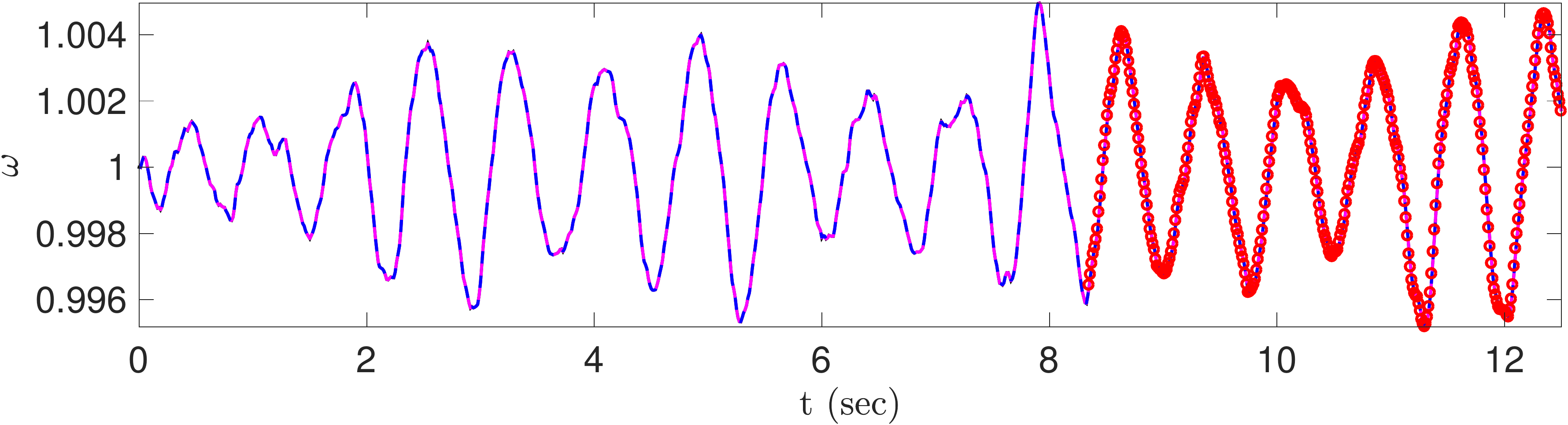}
        \includegraphics[scale=.22]{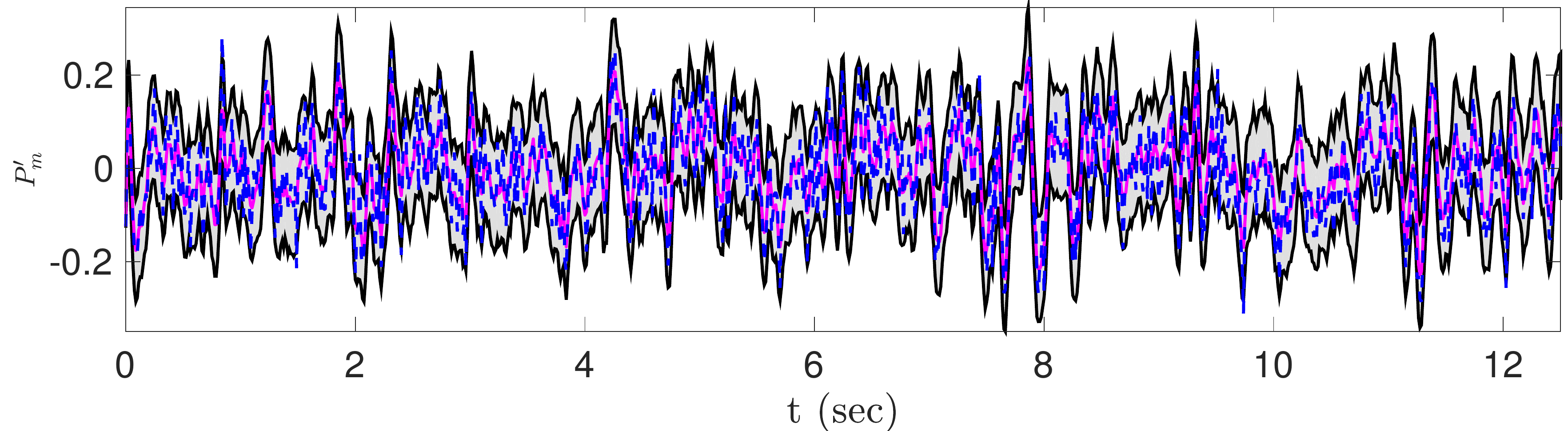}
   \caption{Forecast of $\theta$, $\omega$, and $P'_m$ with $333$  observations of $\omega$ for $t > 8.3375$. } \label{theta_omega_sample_10_obs_omega_333}
\end{figure}
This case is similar to Case 2, except  instead of $\theta(t)$ measurements, we have $\omega(t)$ measurements for  $t <8.3375$.  Our objective is to predict $\omega(t)$ for $t > 8.3375$ and $\theta(t)$ and $P'_m(t)$ for the entire time interval $t\in[0,12.5]$ s. 
Fig. \ref{theta_omega_sample_10_obs_omega_0} presents measurements of $\omega$ and predicted  $\omega(t)$, $\theta(t)$, and $P'_m(t)$, as well as the ground truth solutions for these variables. As in Case 2, physics-informed GPR provides excellent predictions of $\theta(t)$ and $P'_m(t)$ for $t <8.3375$ and a similarly good prediction for $t > 8.3375$ s as in Case 1 where the $\theta(t)$ measurements  for $t <8.3375$ are available.

Next, we assume that 333 $\omega(t)$ measurements (shown by red circles in Fig. \ref{theta_omega_sample_10_obs_omega_333}) are available for $t > 8.3375$. Fig. \ref{theta_omega_sample_10_obs_omega_333} shows that these measurements significantly improve prediction of the (unobserved) $\theta(t)$ and $P'_m(t)$ variables. 

\begin{figure}[h!]
    \centering   
        \includegraphics[scale=.22]{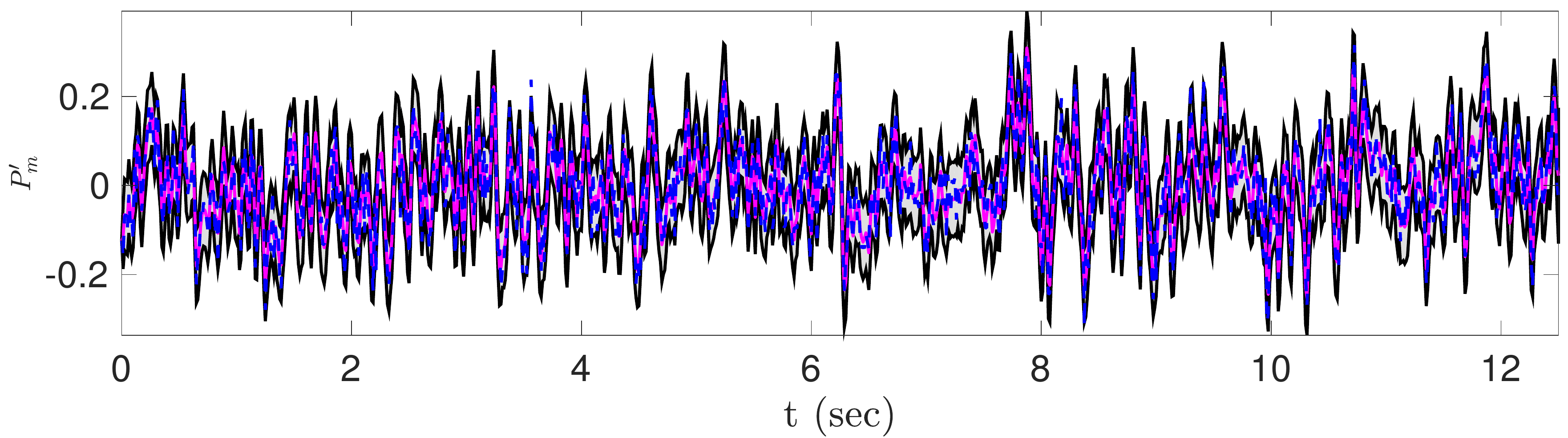}
        \includegraphics[scale=.22]{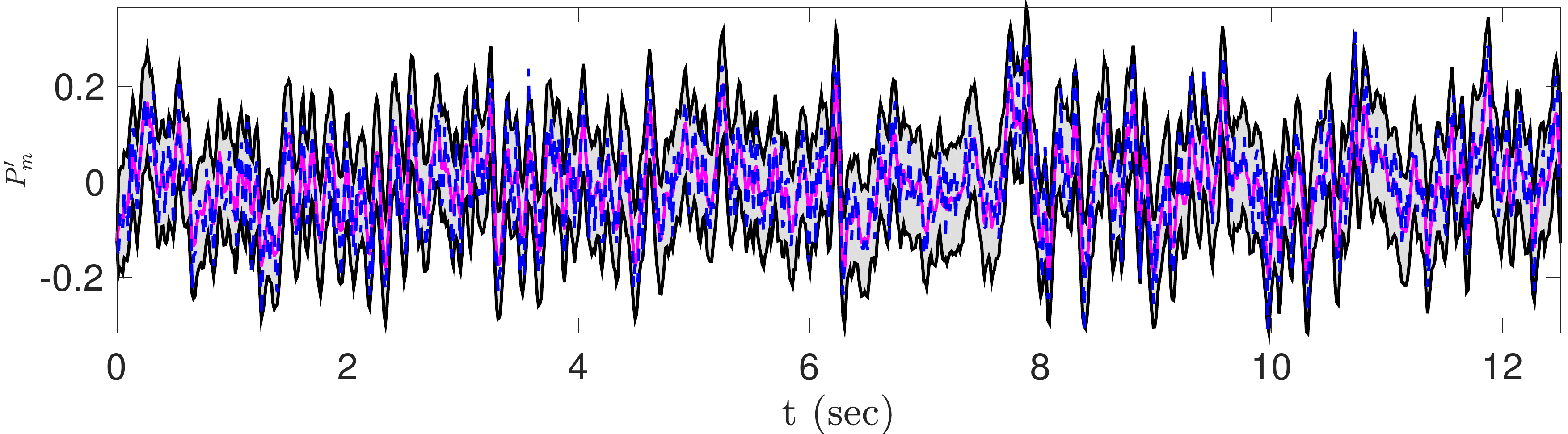}
\caption{Forecast of $P'_m$ process when observations of a) $\theta$ and b) $\omega$ are available.} \label{theta_omega_sample_10_ou_obs_theta_omega_333}
\end{figure}

Fig. \ref{theta_omega_sample_10_ou_obs_theta_omega_333} depicts a comparison of the physics-informed GPR prediction of $P'_m$ based on measurements of $\theta$ and $\omega$.  This comparison reveals that the predictions of $P'_m(t)$  are reasonably good in both cases and better when $\theta$ measurements are available instead of  $\omega$ measurements. This shows that $\theta$ is correlated more strongly with $P'_m$ than $\omega$.

\section{Discussion and conclusions}

We have presented a physics-informed Gaussian Process Regression (GPR) model to predict the phase angle, angular speed, and wind mechanical power when past measurements of all or some of these states are available. In the traditional data-driven GPR method,  a form of the covariance matrix of the Gaussian Process model is first assumed and then its parameters are found from measurements. In the physics-informed GPR, we treat unknown variables, including  wind speed and mechanical power, as a random process and compute the covariance matrix of the GPR model from the resulting stochastic power grid equations. We have demonstrated that the physics-informed GPR method is significantly more accurate than the standard data-driven GPR method for forecasting generators' angular velocity and phase angle. We also have shown that the physics-informed GPR provides accurate predictions of the unobserved wind mechanical power, phase angle, or angular velocity when measurements of only one of these variables are available. 

One advantage of GPR over neural-network-based methods is that its predictions come with uncertainty bounds, i.e., GPR predicts the mean value and standard deviation of quantities of interest. We have demonstrated that the physics-informed GPR accurately and reliably (with small uncertainty) predicts unobserved variables from observed ones for the entire time interval where observations of other (correlated) variables are available. For forecasting, the uncertainty in predicted values slowly increases with the time horizon. In the considered cases, the  forecasted mean values were accurate in the time horizon equal to as many as 200 correlation times of the wind power fluctuations. For larger time horizons, we have noted the predicted values to be within two standard deviations from the ground truth. 

In this study, we have modeled high-frequency power grid dynamics produced by high-frequency wind oscillations. The correlation length of the wind mechanical power in the simulated data was $\lambda = 0.026$ s, and the accurate mean values were obtained for up to 2 s time horizon. Such high-frequency oscillations have been shown to have a significant effect on small signal stability   \cite
{Rosenthal2017TPS,Wang2015JUQ,Barajas2016PRE}. For economic load dispatch and planning, the considered wind oscillations are on the scale of minutes. With $\lambda=2$ min, we expect the physics-informed GPR to provide an accurate forecast for the 6 hr time horizon, which is sufficient for load dispatch and planning  \cite{soman2010review}.  Unit commitment, reserve requirement decisions, and maintenance scheduling require 24 hrs to one-week forecasts \cite{soman2010review}. In our future work, we will investigate if the physics-informed GPR can provide an accurate forecast on such time horizons by considering low-frequency wind and load fluctuations with the correlation time on the order of 0.1 - 0.5 hrs.


\addtolength{\textheight}{-.2cm} 

\section{Acknowledgment}
This work was partially
supported by the U.S. Department of Energy (DOE) Office of
Science, Office of Advanced Scientific Computing Research
(ASCR) as part of the Multifaceted Mathematics for Rare,
Extreme Events in Complex Energy and Environment Systems
(MACSER) project. Pacific Northwest National Laboratory
 is operated by Battelle for the DOE
under Contract DE-AC05-76RL01830.



\end{document}